\documentclass[sigconf]{acmart}
\AtBeginDocument{%
  \providecommand\BibTeX{{%
    \normalfont B\kern-0.5em{\scshape i\kern-0.25em b}\kern-0.8em\TeX}}}

\setcopyright{acmcopyright}
\copyrightyear{2023}
\acmYear{2023}
\acmDOI{XXXXXXX.XXXXX}

\acmConference[Conference acronym 'XX]{Make sure to enter the correct
  conference title from your rights confirmation emai}{June 03--05,
  2023}{Woodstock, NY}
\acmISBN{978-1-4503-XXXX-X/18/06}

\usepackage{multirow}
\usepackage{xcolor}
\usepackage{paralist}

\begin{document}

% \title{Selective Data Replay} 
% \title{How to Predict CVR Accurately in Sales Promotions? A Novel CVR Prediction Framework to Handle Distribution Saltation}

% \title{How to Resolve CVR Fluctuation during Sales Promotion? \\ A Novel Data Reuse Approach for CVR Estimation} 
% \title{How to Address CVR Fluctuation during Sales Promotion? \\ A Novel CVR Estimation Approach via Data Reuse} 
\title{COPR: Consistency-Oriented Pre-Ranking for Online Advertising} 
% \title{\chan{Calibrated} Conversion Rate Estimation during Sales Promotion} 
\renewcommand{\shorttitle}{COPR: Consistency-Oriented Pre-Ranking for Online Advertising} 


% \author{Zhangming Chan, Shuguang Han\textsuperscript{\small{$\ast$}}, Yu Zhang, Yong Bai\textsuperscript{\small{$\dagger$}}, Xiang-Rong Sheng \\ Siyuan Lou, Jiacen Hu\textsuperscript{\small{$\ddagger$}}, Baolin Liu\textsuperscript{\small{$\ddagger,\ast$}}, Yuning Jiang, Jian Xu, Bo Zheng} 
% \author{Zhangming Chan, Yu Zhang, Shuguang Han*\authornote{Shuguang Han and Baolin Liu are the corresponding authors.}, Yong Bai\textsuperscript{\small{$\dagger$}}, Xiang-Rong Sheng, \\ Siyuan Lou, Jiacen Hu\textsuperscript{\small{$\ddagger$}}, Baolin Liu\textsuperscript{\small{$\ddagger$}}, Yuning Jiang, Jian Xu, Bo Zheng} 
% \affiliation{
%     \institution{Alibaba Group \textsuperscript{\small{\quad $\dagger$ }}Nanjing University \textsuperscript{\small{\quad $\ddagger$ }}University of Science and Technology Beijing} 
%     \city{Beijing \& Nanjing} 
%     % \country{People's Republic of China} 
%     \country{China} 
% }
% % \authornote{Shuguang Han and Baolin Liu are the corresponding authors.}
% \email{{zhangming.czm,xieyuan.zy,shuguang.sh,xiangrong.sxr,lousiyuan.lsy,mengzhu.jyn}@alibaba-inc.com} 
% \email{baiy@smail.nju.edu.cn, {hujiacen,liubaolin}@ustb.edu.cn, {xiyu.xj,bozheng}@alibaba-inc.com} 
% \email{{zhangming.czm,shuguang.sh}@alibaba-inc.com, liubaolin@ustb.edu.cn}

\author{Zhishan Zhao, Jingyue Gao}
\authornote{Zhishan Zhao and Jingyue Gao contribute equally to this work.}
% \email{{zhaozhishan.zzs,jingyue.gjy}@alibaba-inc.com}
\affiliation{%
  \institution{Alibaba Group}
    \city{Beijing}
    \country{China}
}

\author{Yu Zhang, Shuguang Han}
% \email{{xieyuan.zy,shuguang.sh}@alibaba-inc.com}
\affiliation{%
  \institution{Alibaba Group}
    \city{Beijing}
    \country{China}
}

\author{Siyuan Lou, Xiang-Rong Sheng}
% \email{{lousiyuan.lsy,xiangrong.sxr}@alibaba-inc.com}
\affiliation{%
  \institution{Alibaba Group}
    \city{Beijing}
    \country{China}
}

\author{Zhe Wang, Han Zhu}\authornote{Han Zhu is the corresponding author.}
% \email{zhuhan.zh@alibaba-inc.com}
\affiliation{%
  \institution{Alibaba Group}
    \city{Beijing}
    \country{China}
}

\author{Yuning Jiang, Jian Xu}
% \email{mengzhu.jyn@alibaba-inc.com}
\affiliation{%
  \institution{Alibaba Group}
    \city{Beijing}
    \country{China}
}

\author{Bo Zheng}
% \email{bozheng@alibaba-inc.com}
\affiliation{%
  \institution{Alibaba Group}
    \city{Beijing}
    \country{China}
}

\renewcommand{\authors}{Zhishan Zhao, Jingyue Gao, Yu Zhang, Shuguang Han, Siyuan Lou, Xiang-Rong Sheng, Zhe Wang, Han Zhu, Yuning Jiang, Jian Xu, Bo Zheng}
\renewcommand{\shortauthors}{Zhao and Gao, et al.}

\begin{abstract}
Cascading architecture has been widely adopted in large-scale advertising systems to balance efficiency and effectiveness. In this architecture, the pre-ranking model is expected to be a lightweight approximation of the ranking model, which handles more candidates with strict latency requirements.  Due to the gap in model capacity, the pre-ranking and ranking models usually generate inconsistent ranked results, thus hurting the overall system effectiveness. The paradigm of score alignment is proposed to regularize their raw scores to be consistent. However, it suffers from inevitable alignment errors and error amplification by bids when applied in online advertising. To this end, we introduce a consistency-oriented pre-ranking framework for online advertising, which employs a chunk-based sampling module and a plug-and-play rank alignment module to explicitly optimize consistency of ECPM-ranked results. A $\Delta NDCG$-based weighting mechanism is adopted to better distinguish the importance of inter-chunk samples in optimization. Both online and offline experiments have validated the superiority of our framework. When deployed in Taobao display advertising system, it achieves an improvement of up to +12.3\% CTR and +5.6\% RPM.  
\end{abstract} 



\maketitle

\section{Introduction}
Online advertising has become a major source of revenue for many web platforms~\cite{goldfarb2011online,Google2022}. Advertisers ensure effective promotion of products by bidding and paying for user actions (e.g., click and purchase)\footnote{Without loss of generality, we regard click as the action in this paper} on advertisements (i.e., ads). To maximize platform revenue, the advertising system typically ranks ads based on their Expected Cost Per Mille (ECPM)~\cite{yuan2019improving} and selects top ones for impression:
\begin{equation}
    ECPM = 1000 \times bid \times pCTR ,
\end{equation}\label{eq:ecpm}where $bid$ is the price that the advertiser is willing to pay and $pCTR$ is the  predicted click-through rate (CTR) denoting the probability that the user clicks the ad.

\begin{figure}[!tbp]
	\centering	
	\includegraphics[width=0.65\columnwidth]{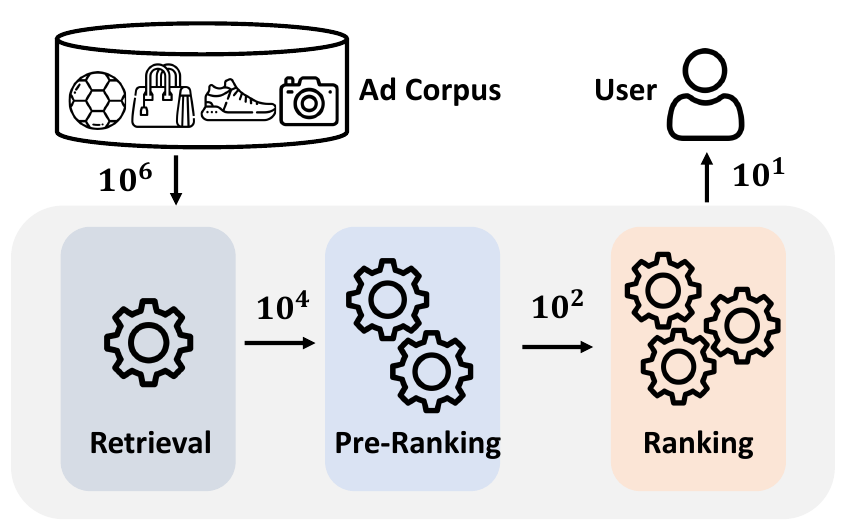}
    \caption{An illustration of the typical three-phase cascading architecture for online advertising systems.}\label{fig:cascading}
\end{figure}

Under strict latency requirements in online deployment, it is infeasible for complex CTR models~\cite{zhou2018deep,liu2020autofis,pi2020search,guo2021embedding} with high inference cost to handle millions of candidates in the ad corpus. To balance
efficiency and effectiveness, a common practice in industrial systems is to adopt a cascading architecture~\cite{covington2016deep,wang2020cold,qin2022rankflow,li2022inttower}, which filters ads through multiple phases with increasingly complex models as illustrated in Fig.~\ref{fig:cascading}. Particularly, the retrieval model first retrieves tens of thousands of relevant ads from the corpus. Afterwards, the pre-ranking model outputs pCTR for retrieved candidates, where top hundreds with highest ECPM are sent to the ranking model for final selection. To handle a larger candidate set, the pre-ranking model is usually designed to be lightweight, which works more efficiently but less accurately compared with the ranking model. 

 Pre-Ranking has recently received increasing attention due to its importance in the cascading architecture. Huang et al.~\cite{huang2013learning} propose a two-tower model that maps users and candidates into latent vectors and calculates their inner products. To enable high-order feature interactions, Li et al.~\cite{li2022inttower} add find-grained interactions between two towers and Wang et al.~\cite{wang2020cold} propose to use deep neural network with squeeze-and-excitation block. Despite improvement of accuracy, there is still a non-negligible gap between the pre-ranking and ranking models. They may generate significantly different ranked results on the same candidate set. Such \textbf{inconsistency} hinders the overall system effectiveness. For example, top ads selected from the pre-ranking phase could be less competitive in the ranking phase, causing waste of the computational resource. Also, ads which are preferred in the ranking phase could be unfortunately discarded in the pre-ranking phase, leading to sub-optimal results. 

Some pioneering studies~\cite{qin2022rankflow,tang2018ranking} propose to align the pre-ranking and ranking models via distillation on pCTR scores. The pre-ranking model is encouraged to generate same scores as the ranking model~\cite{qin2022rankflow} or generate high scores for top candidates selected by the ranking model~\cite{tang2018ranking}.  Although exhibiting encouraging performance, the paradigm of \textbf{score alignment} suffers from the following issues, especially when applied to the advertising system:
\begin{itemize}
    \item \textbf{Inevitable alignment errors.} Due to simpler architecture and fewer parameters for efficiency concerns, the capacity of the pre-ranking model is limited, making it difficult to well approximate original scores of the complex ranking model. Thus even with explicit optimization, there still exist errors in aligning their scores to be exactly the same.
    \item \textbf{Error amplification in ECPM ranks\footnote{We use \textbf{ECPM rank} to denote the order of an ad in the ECPM-ranked list.}.} In both pre-ranking and ranking phases, ads are ranked according to their ECPM as Eq.~(\ref{eq:ecpm}), which is jointly determined by the pCTR score and the bid. Thus the influence of alignment errors could be amplified due to existence of bids. As shown in Table~\ref{tab:toy}, when multiplied by corresponding bids, even a tiny difference in pCTR scores of the pre-ranking and ranking models leads to completely different ranked results.
\end{itemize}

\begin{table}[!htbp]
	\centering
	\caption{A toy example of error amplification in ECPM ranks. Though pCTR scores of two phases are similar, their ranked results after considering bids are different.}\label{tab:toy}
	\resizebox{.9\columnwidth}{!}{
		\begin{tabular}{c|cccc|cccc}
			\toprule
   \multirow{2}{*}{Candidates} & \multicolumn{4}{c|}{Pre-Ranking} & \multicolumn{4}{c}{Ranking} \\
   \cline{2-9}
			 & bid & pCTR & ECPM & rank & bid & pCTR &  ECPM & rank\\
			\midrule
   A & 21 & 0.1 & 2.1 & 2 & 21 & 0.11 & 2.31 & 1\\
   B & 11 & 0.2 & 2.2 & 1 & 11 & 0.19 & 2.09 & 2\\
			\bottomrule
		\end{tabular}
	}
\end{table}

Above issues call for rethinking the necessity of strictly aligning pCTR scores in the advertising system. Essentially, given a set of candidates, it is \textbf{not their absolute pCTR scores but their relative ECPM ranks} that determine the results of each phase. Therefore, 
to achieve consistent results, the pre-ranking model is not required to output same pCTR scores as the ranking model. Instead, it only needs to output scores which yield same ECPM ranks when multiplied by bids. In this way, the requirement of score alignment can be relaxed to that of \textbf{rank alignment}, which is more easier to meet. Moreover, when optimizing pCTR scores for consistent ECPM ranks, the influence of bids can be taken into account beforehand, thus alleviating the issue of error amplification.  

To this end, we introduce a \underline{\textbf{C}}onsistency-\underline{\textbf{O}}riented \underline{\textbf{P}}re-\underline{\textbf{R}}anking (\textbf{COPR}) framework for online advertising, which explicitly optimize the pre-ranking model towards consistency with the ranking model. Particularly, we collect historical logs of the ranking phase, where each log records a ECPM-ranked list of candidates. COPR segments the list into fixed-sized chunks. Each chunk is endowed with certain level of priority from the view of the ranking phase.  With pairs of ads sampled from different chunks, COPR learns an plug-and-play rank alignment module which aims to consistently distinguish their priority using scores at the pre-ranking phase. Moreover, we adopts a $\Delta NDCG$-based weighting mechanism to better distinguish the importance of inter-chunk pairs in optimization.

Our main contributions can be summarized as follows:
\begin{itemize}
    \item To the best of our knowledge, we are the first to explicitly optimize the pre-ranking model towards consistency with the ranking model in the widely-used cascading architecture for online advertising. 
    \item We propose a novel consistency-oriented pre-ranking framework named COPR, which employs a chunk-based sampling module and a plug-and-play rank alignment module for effective improvement of consistency.
    \item We conduct extensive experiments on public and industrial datasets. Both offline and online results validate that the proposed COPR framework significantly outperforms state-of-the-art baselines. When deployed in Taobao display advertising system, it achieves an improvement of up to +12.3\% CTR and +5.6\% RPM.   
\end{itemize}
\section{Related Work}
In this section, we briefly review studies about pre-ranking.

 Located in the middle of the cascading architecture, the pre-ranking system has played an indispensable role for many large-scale industrial systems~\cite{ma2021towards,wang2020cold}. The development of a pre-ranking model is mainly for balancing the system effectiveness and efficiency, as the downstream ranking model usually cannot deal with tens of thousands of candidates. To this end, techniques such as the dual-tower modeling~\cite{huang2013learning,wu2018eenmf} are commonly adopted. However, this paradigm limits feature interactions between users and items to the form of vector product, which often results in extensive performance degradation.

Another line of work strives to enhance high-order feature interactions, and explores the ways to reduce the online latency. Li et al.~\cite{li2022inttower} add fine-grained and early feature interactions between two towers. Wang et al.~\cite{wang2020cold} propose to use fully-connected layers and employ various techniques from the perspectives of both modeling efficiency and engineering optimization. Specifically, a Squeeze-and-Excitation module~\cite{hu2018squeeze} is utilized to choose the most useful feature set, and meanwhile system parallelism and low-precision computation are exploited whenever possible for latency optimization. Ma et al.~\cite{ma2021towards} propose a feature selection algorithm based on feature complexity and variational dropout (FSCD) to search a set of effective and efficient features for pre-ranking. A similar study~\cite{li2022autofas} uses network architecture searching (NAS) to determine the optimal set of features and corresponding architectures. These studies mainly focus on improving the accuracy of the pre-ranking model but neglects its interaction with the subsequent ranking model, leading to inconsistent ranked results. 

\begin{figure*}[!htbp]
    \centering
    \includegraphics[width=0.88\textwidth]{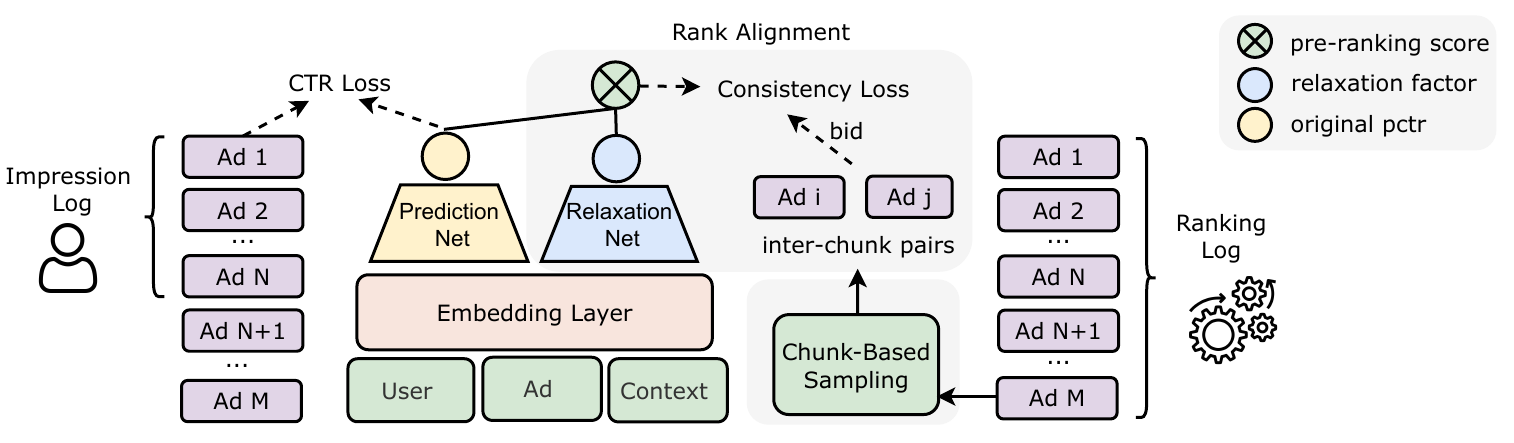}
    \caption{The framework of consistency-oriented pre-ranking.}
    \label{fig:framework}
\end{figure*}

Several studies propose to align the pre-ranking and ranking models in terms of pCTR scores via knowledge distillation. RD~\cite{tang2018ranking} encourages the lightweight student model to score higher for candidates selected by the larger teacher model, which is often used in training pre-ranking models. RankFlow~\cite{qin2022rankflow} regularizes the pre-ranking and ranking models to generate same scores for same candidates. Despite encouraging performance, there still exist inevitable errors in score alignment due to discrepancy in model capacity. When applied in online advertising, influence of such errors would be amplified by bids of ads, yielding inconsistent ECPM-ranked results. In this paper, we propose to relax the objective of
score alignment to rank alignment, where bids of ads are incorporated and consistency of ranked results between two phases can be explicitly optimized in an effective manner.

\section{Methodology}
In this section, we first introduce background knowledge about the pre-ranking model, and then describe our proposed COPR framework as illustrated in Fig.~\ref{fig:framework}.
\subsection{Background}
\noindent\textbf{Training Data.} When the advertising system serves online traffic as Fig.~\ref{fig:cascading}, hundreds of ads are ranked through the ranking phase and recorded to logs, which we refer to as \textbf{ranking logs}. Each log contains an ranked list of ads with descending ECPM: 
\begin{equation}\label{eq:rankinglogs}
    \mathbf{R} = [(ad_1, pCTR_1, bid_1),...,(ad_M, pCTR_M, bid_M)],
\end{equation}
where $pCTR_i$ is the score output by the ranking model for $i$-th ad and $bid_i$ denotes its bid. $M$ is the number of candidates. Then top $N$ ads are displayed to the user. User feedback $y$ (click/non-click) on each displayed ad is recorded to \textbf{impression logs}:
\begin{equation}
    \mathbf{I} = [(ad_1,y_1),...,(ad_N, y_N)].
\end{equation}

\noindent\textbf{Base Model.} The base model for pre-ranking is usually a lightweight CTR model. Here we adopt the architecture of COLD~\cite{wang2020cold}. The input features consist of three parts: user features $\mathbf{U}$ such as age and gender, ad features $\mathbf{A}$ such as brand and category, context features $\mathbf{C}$ such as time and device. After pre-selecting a concise set of features, COLD feeds them into embedding layers and concatenate their embeddings for a compact representation $\mathbf{x}$:
\begin{equation}
    \mathbf{x} = E(\mathbf{U}) \oplus E(\mathbf{A}) \oplus E(\mathbf{C}).
\end{equation}

Then it employs a prediction net consists of multiple fully-connected layers to estimate CTR:
\begin{equation}\label{eq:org_pCTR}
    \hat{y} = Sigmoid(MLP(\mathbf{x})) \in [0,1].
\end{equation}

To accurately predict user click $y$, the model is optimized with cross entropy loss over impression logs $I$:
\begin{equation}
       L_{ctr} = \sum_{\mathbf{I}}[-ylog(\hat{y})-(1-y)log(1-\hat{y})].  
\end{equation}

\subsection{Consistency-Oriented Pre-Ranking}
Though the pre-ranking model is expected to well approximates the ranking model in the cascading system,  their gap in model capacity often hinders satisfying approximation. Thus in addition to $L_{ctr}$, we aim to explicitly optimize the pre-ranking model towards consistent results with the ranking model over $\mathbf{R}$. 

\subsubsection{Chunk-Based Sampling}

\begin{figure}[!htbp]
    \centering
    \includegraphics[width=0.3\textwidth]{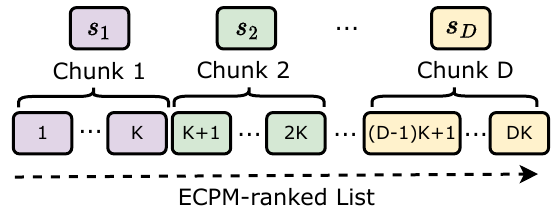}
    \caption{Illustration of chunk-based sampling.}
    \label{fig:chunk}
\end{figure}

Given candidates $\{Ad_i\}_{1}^{M}$ in ranking logs, an ideal pre-ranking model should output scores that yield same ECPM-ranked list as Eq.~(\ref{eq:rankinglogs}). Considering its limited capacity, it could be hard to rank hundreds of ad all in correct positions. To reduce the learning difficulty, we partition the ranked list into $D=\frac{M}{K}$ fixed-sized chunks, each constituting $K$ adjacent ads, as shown in Fig.~\ref{fig:chunk}. We regard ads in the same chunk as candidates with same priority in the ranking phase. The pre-ranking model is not required to distinguish ads in the same chunk. Instead, it only needs to consistently rank candidates in the granularity of chunk. For each chunk, we randomly sample a candidate and endow it with the priority related to this chunk. In this way, for each ranked list, we obtain a concise sub-list:
\begin{equation}\label{eq:chunklogs}
    \mathbf{R}_{chunk} = [(ad_{s_d}, pCTR_{s_d}, bid_{s_d}, D-d)]_{d=1}^{D}, 
\end{equation}
where $s_d$ is the index of sampled ad in chunk $d$ and $D-d$ denotes its priority which the larger the better. 

The above chunk-based sampling has two-fold advantages: 1) It provides a flexible way to control the granularity of consistency, which makes the objective reachable for the lightweight pre-ranking model. By increasing the chunk size $K$,  the objective of consistency gradually shifts from fine-grained to coarse-grained. 2) It effectively reduces the size of ranked list in logs by $K$ times and still maintains coverage of original lists, which is critical for efficient training in industrial machine learning systems. In our production implementation, $K$ is set to 10.

\subsubsection{Rank Alignment} In the following, we introduce how to modify the base model with a plug-and-play rank alignment module. 

Instead of regularizing the difference between $\hat{y}_i$ in Eq.~(\ref{eq:org_pCTR}) and $pCTR_i$ in Eq.~(\ref{eq:chunklogs}) as score alignment methods~\cite{qin2022rankflow,tang2018ranking}, we propose to relax the objective to rank alignment on a properly-adjusted pCTR score. Particularly, we employ a relaxation net to learn a factor $\alpha > 0$, with which we adjust the original pCTR score:

\begin{equation}
    \begin{aligned}
    \alpha &= ReLU(MLP(x))+1e^{-6} \in \mathcal{R^{+}}, \\
    \tilde{y} &= \alpha * \hat{y},
\end{aligned}
\end{equation}
where $\tilde{y}$ denote the adjusted pCTR. Thus ECPM at the pre-ranking phase can be accordingly estimated as $\tilde{y} * bid$, based on which we aim to correctly rank each inter-chunk pair in $\mathbf{R}_{chunk}$. Here we adopt the pairwise logistic loss for its relatively good performance and the simplicity for implementation~\cite{burges2005learning,pasumarthi2019tf}:
\begin{equation}\label{eq:rank}
    L_{rank} = \sum_{i<j}log[1+e^{-(\frac{\tilde{y}_{s_i} * bid_{s_i}}{\tilde{y}_{s_j} * bid_{s_j}}-1) }].
\end{equation}
For each pair of $ad_{s_i}$ and $ad_{s_j}$ sampled from different chunks that $i<j$, we optimize $L_{rank}$ by encouraging $\tilde{y}_{s_i} * bid_{s_i} > \tilde{y}_{s_j} * bid_{s_j}$, which means $ad_{s_i}$ would be ranked before $ad_{s_j}$
by ECPM in the pre-ranking phase. If all inter-chunk pairs can be correctly ranked, we achieve consistent ECPM-ranked results between the pre-ranking and ranking phases over $R_{chunk}$. 

Note that by introducing the relaxation factor $\alpha$, we slightly modify the original pCTR score to achieve consistent ranked results if necessary. To maintain original value as much as possible, $\alpha$ should be around $1$. Thus we add a symmetric regularization to penalize the deviation of $\alpha$ from 1:
\begin{equation}
    L_{reg} = 
    \begin{cases}
        \alpha-1& \alpha>1\\
        \frac{1}{\alpha}-1& \alpha<=1
    \end{cases}.
\end{equation}

It is worth mentioning that the proposed rank alignment module does not rely on specific assumption about the architecture of base model.
It is an plug-and-play component that can be added to any pre-ranking models for improvement of consistency.

\subsubsection{$\Delta NDCG$-Based Pair Weighting}  $L_{rank}$ in Eq.~(\ref{eq:rank}) fails to consider the relative importance of  different pairs in consistency optimization. In practice, consistently ranking ads from chunk $1$ and chunk $10$ is more important than ranking chunk $11$ and chunk $20$, since only the top ads will be sent to the ranking phase and displayed to users. It calls for a weighting mechanism that considers chunk-related priorities of candidates.

Intuitively, if pair $(ad_{s_i}, ad_{s_j})$ in $L_{rank}$ are mistakenly ranked, the consistency between the pre-ranking and ranking phase will be hurt. Thus its weight in $L_{rank}$ should be determined by the negative impact. As each sampled $ad_{s_d}$ in $\mathbf{R}_{chunk}$ is endowed with priority $D-d$, we use NDCG~\cite{jarvelin2017ir,burges2010ranknet} to measure the utility of any ranked list $p$ of these candidates:
\begin{equation}
\begin{aligned}
    DCG &= \sum_{i=1}^{D}\frac{2^{p_i}-1}{log(i+1)},\\
    IDCG &= \sum_{i=1}^{D}\frac{2^{D-i}-1}{log(i+1)},
\end{aligned}
\end{equation}
where $p_i$ denote the priority of $i$-th ad in the permutation and the IDCG is the ideal DCG achieved by $\mathbf{R}_{chunk}$.
If we swap the position of $ad_{s_i}$ and $ad_{s_j}$ in $\mathbf{R}_{chunk}$, the utility of the list will experience a drop which can be further normalized as:
\begin{equation}
    \Delta NDCG(i,j) =  \frac{2^{D-i}-2^{D-j}}{IDCG}[\frac{1}{log(i+1)}-\frac{1}{log(j+1)}].
\end{equation}
The utility drop is used to re-weight inter-chunk pairs in consistency optimization:
\begin{equation}
    L_{rank} = \sum_{i<j}\Delta NDCG(i,j) log[1+e^{-(\frac{\tilde{y}_{s_i} * bid_{s_i}}{\tilde{y}_{s_j} * bid_{s_j}}-1)}].
\end{equation}
Thus the objective function of COPR can be formulated as:
\begin{equation}\label{eq:finalobj}
    L = \underbrace{L_{ctr}}_{CTR\ Loss}+ \underbrace{\lambda_{1}L_{rank} + \lambda_2 L_{reg}}_{Consistency\ Loss},
\end{equation}
where $\lambda_{1}>0$, $\lambda_{2}>0$ are weights for corresponding loss terms. By minimizing $L$, we explicitly optimize the pre-ranking model towards consistency with the ranking model via a plug-and-play rank alignment module. 

\subsection{System Deployment}

\begin{figure}[!htbp]
    \centering
    \includegraphics[width=0.45\textwidth]{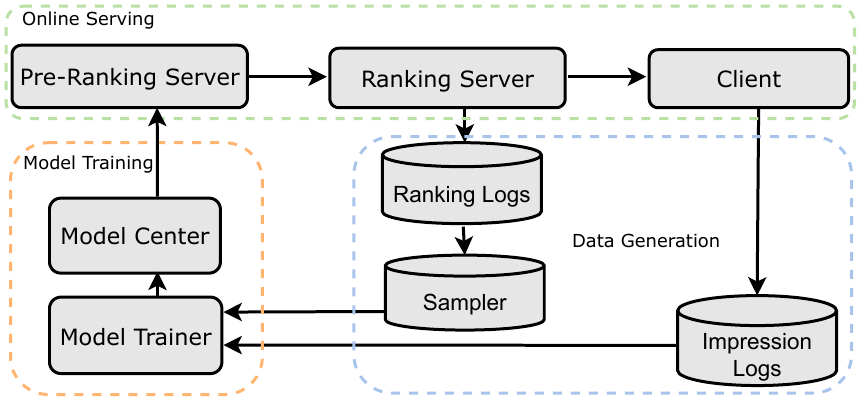}
    \caption{Overview of system pipeline.}
    \label{fig:implementation}
\end{figure}

We introduce the deployment of COPR in three stages: data generation, model training, and online serving as shown in Fig.~\ref{fig:implementation}. 

\noindent\textbf{Data Generation.} During online serving, hundreds of ads are ranked through ranking model and recorded to ranking logs, with which we perform chunk-based sampling. The content of each sample includes user index, ad index, chunk index as well as the bid. Note that the bid at the ranking phase could differ from that at the the pre-ranking phase~\cite{wang2022designing}. In this case, we record the pre-ranking bid since it influences $L_{rank}$ in model training. When ads are displayed to users in the client, we also record user feedback in impression logs, which are used in calculating $L_{ctr}$. 
    
\noindent \textbf{Model Training.} The training procedure is performed on our ODL (Online Deep Learning)~\cite{luoBernoulli} platform, which consumes real-time streaming data to continuously update model parameters. After training with fixed number of steps, the learnt model will be delivered to the Model Center, which manages all online models.

\noindent \textbf{Online Serving.} Once a new version of pre-ranking model is ready, pre-ranking server will load it from Model Center to replace the online version in service.

\section{Experiments}
In this section, we conduct experiments on both public dataset and production dataset to validate the effectiveness of COPR in improving consistency and overall system performance.
\subsection{Experiment Setup} 
\noindent\textbf{Taobao Dataset.} It is a public dataset\footnote{https://tianchi.aliyun.com/dataset/dataDetail?dataId=56} with 26 million impression logs of 1 million users and 0.8 million items in 8 days. Item price is used as bid. Impressions of first 7 days are used to train DIN~\cite{zhou2018deep} as the ranking model. For each impression, we sample 10 candidates and collect ECPM-ranked results by the ranking model to train pre-ranking models. Logs of the last day are used for evaluation. To simulate the cascading process, we sample 100 candidates for each impression, among which the pre-ranking and ranking model sequentially select top 10 and top 1 candidates to display.

\noindent\textbf{Production Dataset.} It contain 8 days of impression logs and ranking logs collected from our system shown in Fig.~\ref{fig:implementation}. These logs are of the magnitude of billions. The first week of logs are used for training and the last day is used for evaluation. According to the scenario that logs come from, it is further divided into two subsets: \textbf{Homepage} and \textbf{Post-Purchase}.

\noindent\textbf{Baselines.} COPR is compared with following baselines:
\begin{compactitem}
    \item \textbf{Base} adopts the architecture of COLD~\cite{wang2020cold} and is trained on impression logs.  
    \item \textbf{Distillation}~\cite{hinton2015distilling} directly distills predicted scores of the ranking model on impression logs. 
    \item \textbf{RankFlow}~\cite{qin2022rankflow} distills predicted scores of the ranking model on ranking logs and further regularizes the pre-ranking model to generate high scores for candidates selected by the ranking model.
    \item \textbf{COPR w/o $\Delta NDCG$} removes the $\Delta NDCG$-based weighting mechanism from the COPR framework.
\end{compactitem}

\noindent\textbf{Metrics.} We adopt two groups of metrics in evaluation. 
\begin{compactitem}
    \item The first group measures the consistency between ECPM-ranked results of the pre-ranking and ranking phases, including HitRatio(\textbf{HR@K}), normalized discounted cumulative gain (\textbf{NDCG@K}), and mean average precision (\textbf{MAP@K}). In HR@K and MAP@K, top 10 candidates selected by the ranking model are treated as relative ones. In NDCG@K, order in  ranking logs is used as a proxy of relevance. The standard calculation of these metrics can be found in~\cite{liu2009learning}.
    \item The second group measures the overall system performance. We use Click-Through-Rate (\textbf{CTR}) and Revenue Per Mille (\textbf{RPM}) similar to~\cite{qin2022rankflow,wang2015real}, which corresponds to user experience and platform revenue, respectively. On  public dataset, CTR is simulated as the portion of clicked ads in displayed ads, and RPM is simulated as the product of CTR and average bid of clicked ads. In production experiment, we perform online A/B test to obtain CTR and RPM on real traffic.     
\end{compactitem}

\noindent\textbf{Hyper-parameters.} The chunk size is set to 2 and 10 on the public dataset and the production dataset, respectively. The number of MLP layers in the prediction net and the relaxation net is 3. The embedding size of raw input features is set to 16. $\lambda_1$ and $\lambda_2$ in Eq.~(\ref{eq:finalobj}) are fixed to 1 and 0.2. 

\subsection{Results on Public Dataset}
\begin{table}[!htbp]
	\centering
	\caption{Comparison among COPR and baselines int terms of consistency and overall system performance. Best results are highlighted in bold.}\label{tab:performance}	
	\resizebox{.98\columnwidth}{!}{
		\begin{tabular}{c|ccc|cc}
			\toprule
			Method & HR@10 & NDCG@10 & MAP@10 & CTR & RPM\\
			\midrule
			Base& 0.544 & 0.402 & 0.198 & 0.0170 & 112.66\\
			Distillation& 0.593 & 0.461 & 0.244 & 0.0176 & 117.36\\
			RankFlow & 0.722 & 0.467 & 0.270 & 0.0182 & 121.36\\
            \midrule
            COPR w/o $\Delta NDCG$& 0.741 & 0.514 & 0.327 & 0.0188 & 129.20\\
            COPR & \textbf{0.759} & \textbf{0.530} & \textbf{0.359} & 0.0191 & 132.93\\
			\bottomrule
		\end{tabular}
	}
\end{table}

Table~\ref{tab:performance} compares COPR and baselines in terms of consistency and system performance. We only show $K=10$ in HR@K, NDCG@K, and MAP@K due to limited space. Results under other settings of $K$ are similar. From Table~\ref{tab:performance}, we draw the following conclusions.

First, system performance (CTR and RPM) is highly associated with the consistency between the pre-ranking and ranking phases. For COPR and baselines, the higher consistency generally yields the better system performance. It validates our motivation to explicitly optimize consistency between phases in order to improve the overall effectiveness of the cascading system.

Second, COPR achieves best consistent results of all methods, outperforming the state-of-the-art RankFlow by 5.1\%, 13.5\%, and 33.0\% in terms of HR@10, NDCG@10, and MAP@10. We attribute the improvement to our shift of objective from score alignment to rank alignment. By such relaxation, COPR can directly optimize towards consistent ECPM-ranked results and meanwhile reduce the learning difficulty for the lightweight model. Moreover, the influence of bids is considered in training COPR, thus alleviating the issue of error amplification that RankFlow suffers from. We also find that RankFlow is better than Distillation. We think it is because Rankflow aligns scores over ranking logs while the latter is on impression logs which is too sparse.

Third, COPR w/o $\Delta NDCG$ experiences performance drop compared with COPR. This ablation study verifies the effectiveness of the pair weighting mechanism based on $\Delta NDCG$. By emphasizing more on important inter-chunk pairs in consistency optimization, COPR ensures top candidates are more likely to be consistently ranked, which helps improve the overall utility of pre-ranking results.

\subsection{Results on Production Dataset} 
\begin{figure}[!htbp]
	\centering	
	\includegraphics[width=0.47\columnwidth]{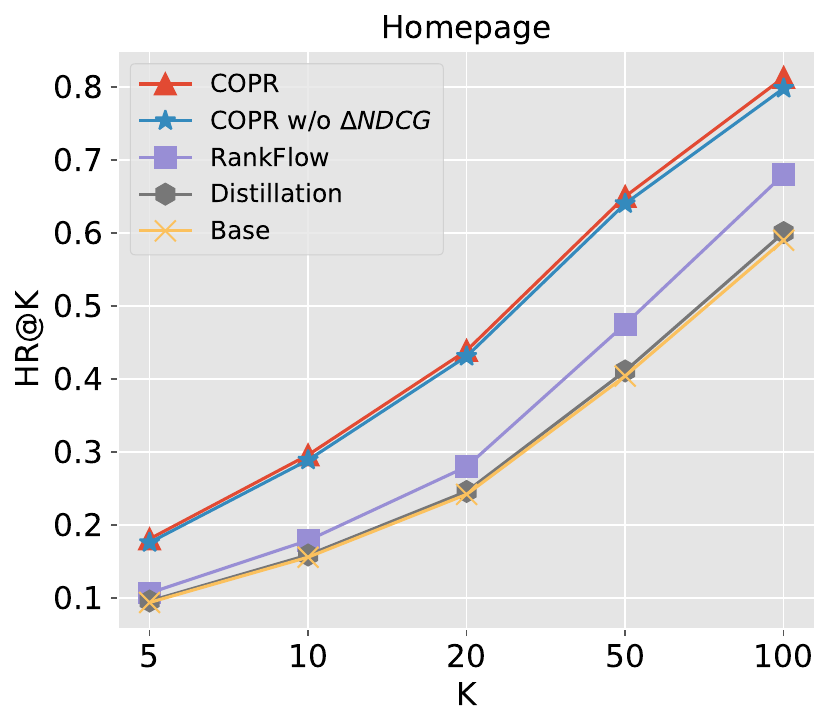}    
	\includegraphics[width=0.47\columnwidth]{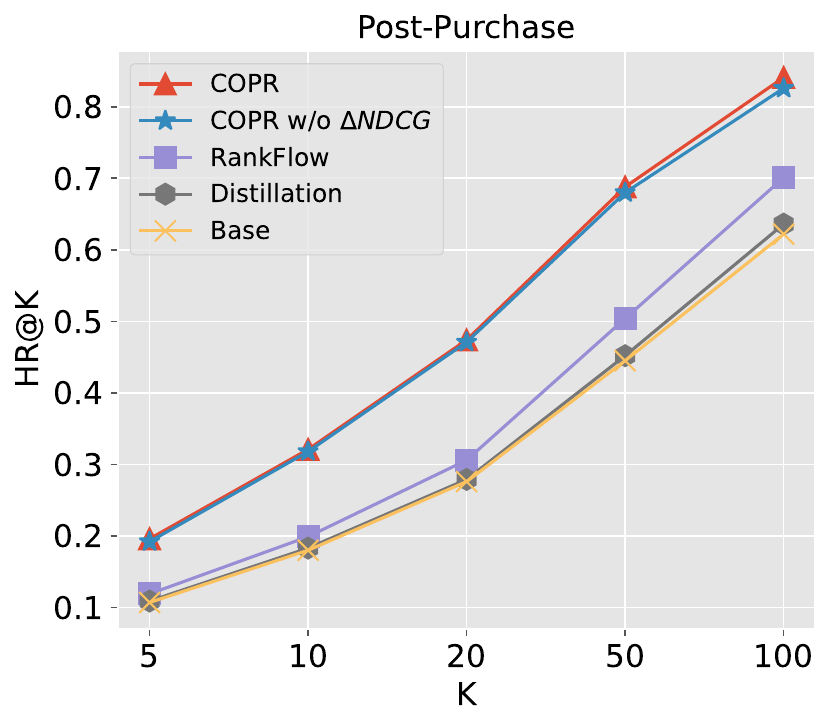}   
	\caption{HR@K of different methods in the scenario of Homepage (Left) and Post-Purchase (Right).}\label{fig:hr}
\end{figure}

\begin{figure}[!htbp]
	\centering	
	\includegraphics[width=0.47\columnwidth]{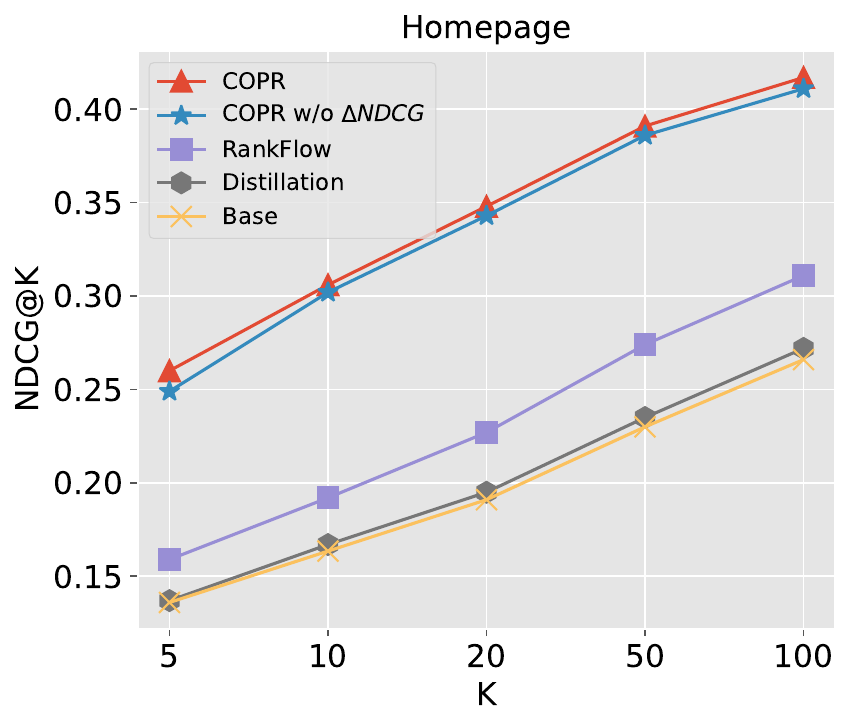}    
	\includegraphics[width=0.47\columnwidth]{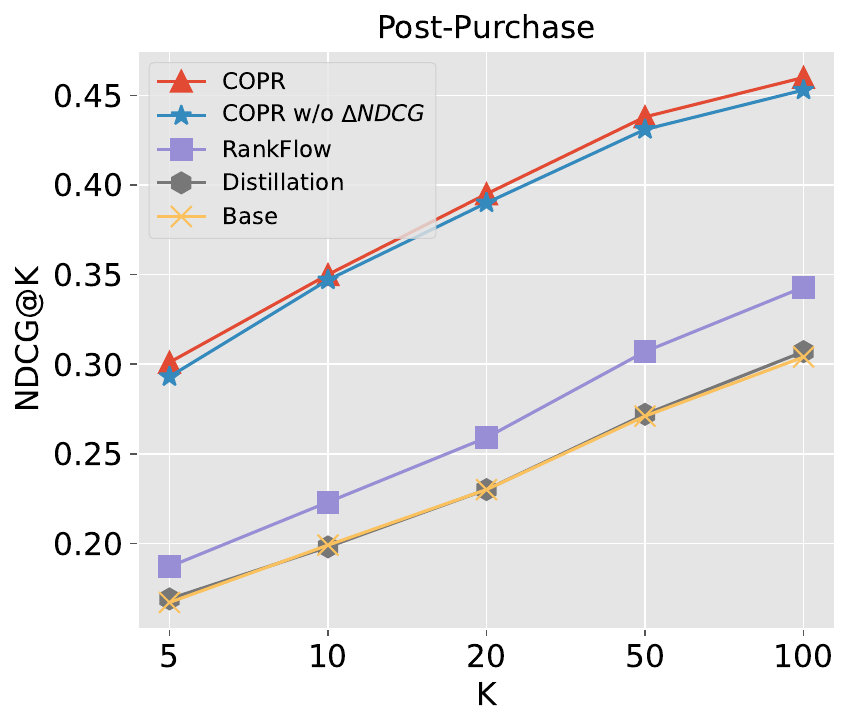}   
	\caption{NDCG@K of different methods in the scenario of Homepage (Left) and Post-Purchase (Right).}\label{fig:ndcg}
\end{figure}

\begin{figure}[!htbp]
	\centering	
	\includegraphics[width=0.47\columnwidth]{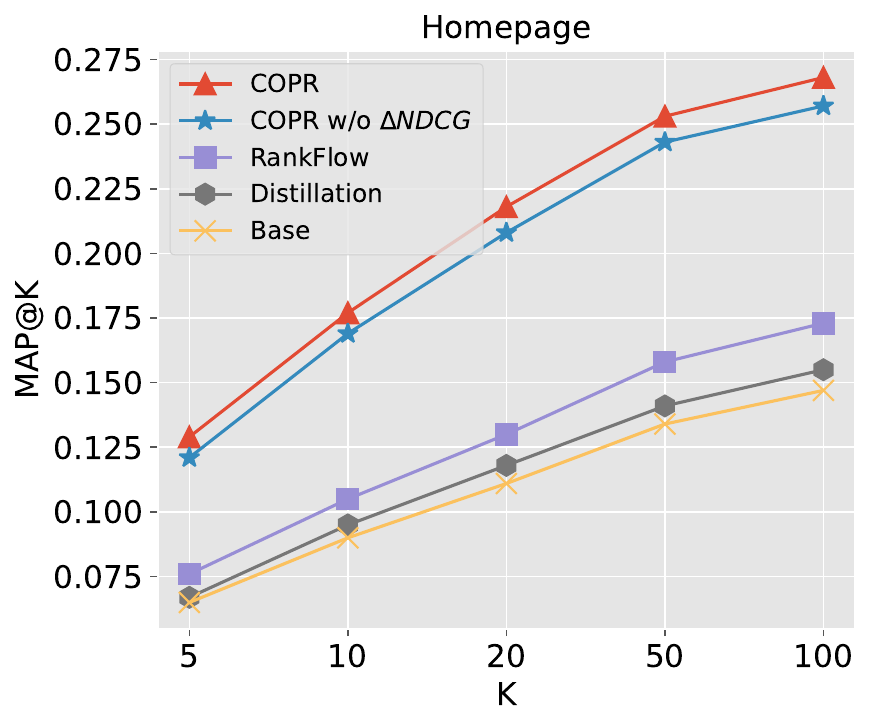}    
	\includegraphics[width=0.47\columnwidth]{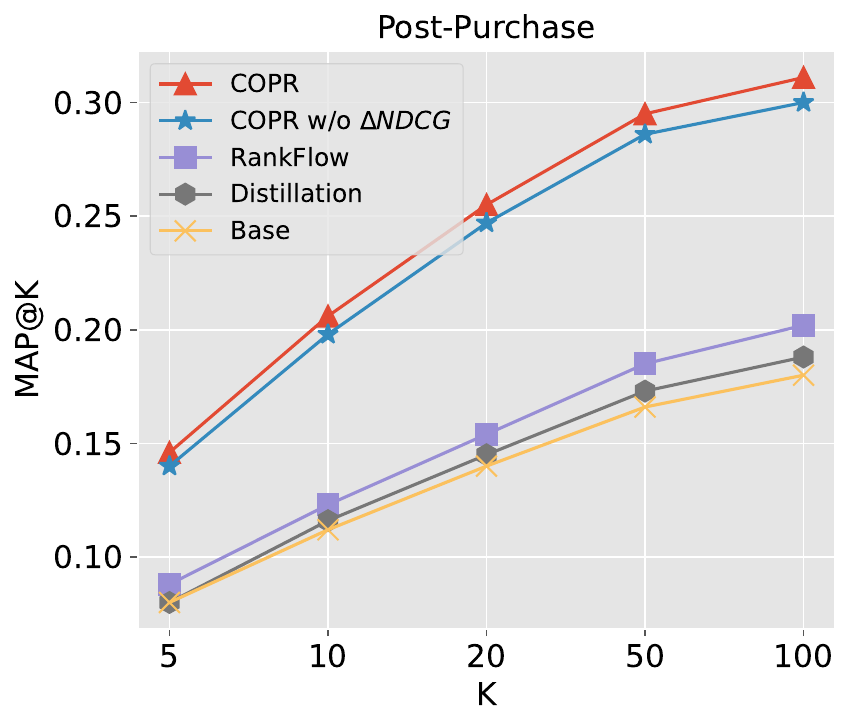}   
	\caption{MAP@K of different methods in the scenario of Homepage (Left) and Post-Purchase (Right).}\label{fig:map}
\end{figure}

We also perform similar evaluation on the production dataset composed of samples from two scenarios. Most conclusions are consistent with those on the public dataset.

As shown in more details from Fig.~\ref{fig:hr} to Fig.~\ref{fig:map}, COPR significantly outperforms other methods in term of HR@K, NDCG@K, and MAP@K with varying $K$ from 5 to 100 on two scenarios, which demonstrates the stable improvement of consistency achieved by our proposed framework. Moreover, we still observe the gap between COPR and COPR w/o $\Delta NDCG$, which shows that the weighting mechanism also works in the large-scale production dataset.

To evaluate system performance in production environment, we perform online A/B test on two scenarios, where these methods are used to serve real users and advertisers. From Table~\ref{tab:online-ab} we find that Distillation, RankFlow, and COPR all perform better than the production baseline, among which COPR achieves the largest improvement, with a lift of up to +12.3\% CTR and +5.6\% RPM. With impressive performance, \textbf{COPR has been successfully deployed to serve the main traffic of Taobao display advertising system in the pre-ranking phase since October of 2022}.

\begin{table}[!htbp]
	\centering
	\caption{Relative improvement over the production baseline in online A/B Test. Best results are highlighted in bold.}\label{tab:online-ab}		
	\resizebox{.75\columnwidth}{!}{
		\begin{tabular}{c|cc|cc}
			\toprule
            \multirow{2}{*}{Method} & \multicolumn{2}{c|}{Homepage} & \multicolumn{2}{c}{Post-Purchase}\\
            \cline{2-5}
            & CTR & RPM & CTR & RPM\\
            \midrule
			Base& -& -& - & -\\
            Distillation & +2.2\% & +0.1\% & +3.6\% & +0.6\%\\
            RankFlow & +8.3\% & +2.9\% & +6.8\% & +2.3\%\\
            \midrule
            COPR w/o $\Delta NDCG$ & +11.5\% & +5.0\% & +9.6\% & +3.7\%\\
            COPR & \textbf{+12.3\%} & \textbf{+5.6\%} & \textbf{+10.8\%} & \textbf{+4.4\%}\\
			\bottomrule
		\end{tabular}
	}
\end{table}

\subsection{Qualitative Analysis}
Given ranked results from the pre-ranking and ranking phases, we calculate the average pre-ranking position for candidates at each ranking position, based on which we draw the Ranking-PreRanking Curve (\textbf{RPC}). The ideal RPC happens when results are exactly same.

\subsubsection{Error Amplification in ECPM Rank.} As shown in Fig.~\ref{fig:quali_consist} (Left), RPC by pCTR of RankFlow is close to the ideal curve, showing well alignment of raw pCTR in two phases. However, after ranking by ECPM, RPC of RankFlow largely deviates from the ideal one. It verifies that the involvement of bid in ECPM will amplify the influence of errors in score alignment, leading to more inconsistent ECPM-ranked results. This analysis is consistent with the example in Table~\ref{tab:toy}. Hence we confirm that merely score alignment is not enough for the cascading architecture in online advertising.

\subsubsection{More Consistent ECPM Rank.} 
\begin{figure}[!h]
	\centering	
	\includegraphics[width=0.47\columnwidth]{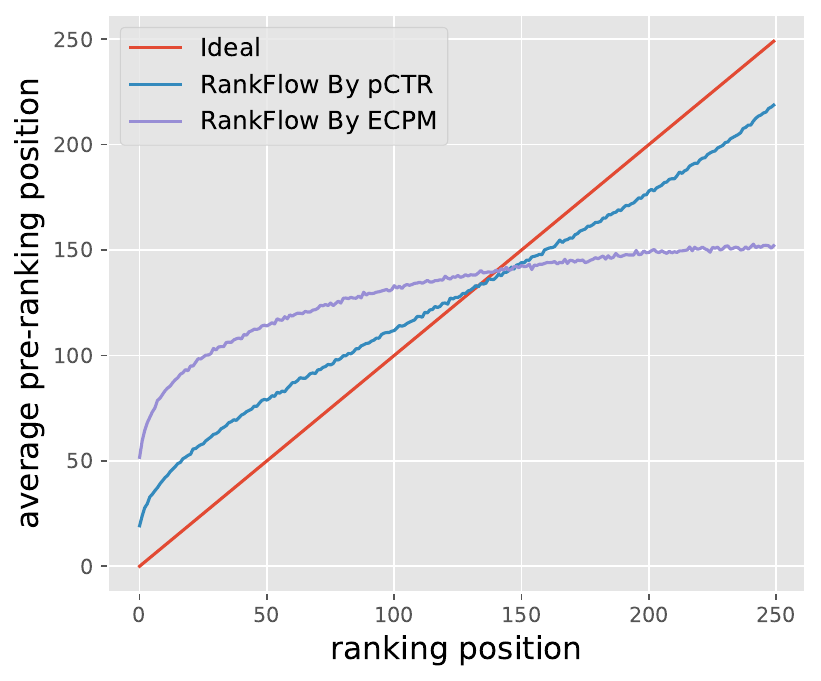}    
	\includegraphics[width=0.47\columnwidth]{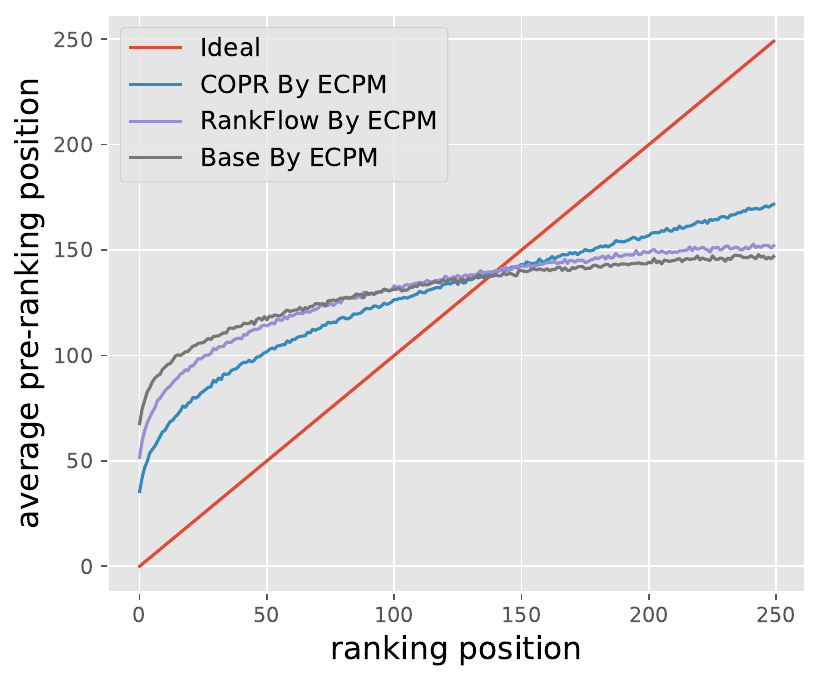}   
	\caption{Left: RPC by pCTR and ECPM of RankFlow. Right: RPC by ECPM of different methods.}\label{fig:quali_consist}
\end{figure}

Fig.~\ref{fig:quali_consist} (Right) shows RPC by ECPM of different methods. We observe that compared with Base and RankFlow, RPC of COPR is more close to the ideal curve in almost each ranking position. It qualitatively shows that ECPM-ranked results given by COPR are more consistent with results of the ranking phase. It can be attributed to the design of our consistency-oriented framework, where the rank alignment module directly optimizes towards this objective. The incorporation of bid also helps alleviate the above mentioned error amplification. 

\section{Conclusion}
In this paper, we introduce a consistency-oriented pre-ranking framework for online advertising, which employs a chunk-based sampling module and a plug-and-play rank alignment module to explicitly optimize consistency of ECPM-ranked results. A $\Delta NDCG$-based weighting mechanism is also adopted to better distinguish the importance of inter-chunk samples in optimization. Both online and offline experiments have validated the superiority of our framework. When deployed in Taobao display advertising system, it achieves an improvement of up to +12.3\% CTR and +5.6\% RPM.

% \section*{Acknowledgements} 
% We would like to thank the anonymous reviewers for their constructive comments. 
% This work was supported by Alibaba Group through Alibaba Research Intern Program, the National Natural Science Foundation of China (No.U2133218), the National Key Research and Development Program of China (No.2018YFB0204304) and the Fundamental Research Funds for the Central Universities of China (No.FRF-MP-19-007 and No. FRF-TP-20-065A1Z). 

\balance 
\bibliographystyle{ACM-Reference-Format}
\bibliography{reference}

\end{document}